\documentclass[twocolumn]{revtex4}
\usepackage{graphicx}
\usepackage{dcolumn}
\usepackage{bm}

\begin{document}
\pagenumbering{arabic}
  % Title of the article
  \title{Model of tunneling transistors based on graphene on SiC}
  % Authors
  \author{Paolo~Michetti, Martina~Cheli, Giuseppe~Iannaccone}
  \affiliation{Dipartimento di Ingegneria dell'Informazione: Elettronica, Informatica, Telecomunicazioni, Universit\`a~ di Pisa, Via Caruso 16, 56122 Pisa, Italy}
  \email{pl.michetti@gmail.com, {martina.cheli, g. iannaccone}@iet.unipi.it}

  \begin{abstract}
Recent experiments shown that graphene epitaxially grown on Silicon Carbide (SiC) can exhibit a energy gap of $0.26$~eV, making it a promising material for electronics.
With an accurate model, we explore the design parameter space for a fully ballistic graphene-on-SiC Tunnel Field-Effect Transistors (TFETs), and assess the DC and high frequency figures of merit.
The steep subthreshold behavior can enable $I_{\rm ON}/I_{\rm OFF}$ ratios exceeding $10^4$ even with a low supply voltage of $0.15$~V, for devices with gatelength down to $30$~nm.
Intrinsic transistor delays smaller than 1~ps are obtained. 
These factors make the device an interesting candidate for low-power nanoelectronics beyond CMOS.

  %We study the performance attainable by graphene-on-SiC Tunnel
  %  Field-Effect Transistors (TFETs) in low-power nanoelectronic systems.
    %
  %  Recent experiments have shown that a two-dimensional layer of
  %  epitaxial graphene on SiC can exhibit a energy gap of $0.26$~eV, making it promising as a material for electronics even without extremely advanced lithography.
    %
   % With an accurate model, we explore the design parameter space for a fully ballistic device, and 
    %assess the DC and high frequency figures of merit.
    %
    %We show that the steep subthreshold behavior can enable $I_{\rm ON}/I_{\rm OFF}$ ratios 
    %exceeding $10^4$ even with a low supply voltage of $0.15$~V, for devices with gatelength down to $30$~nm.
    %
    %Intrinsic transistor delays smaller than 1~ps are obtained. 
    %
    %These factors make the device a very interesting candidate for extreme low-power nanoelectronics beyond CMOS. 
  \end{abstract}
  
  \maketitle
  
  %%%%%%%%%%%%%%%%%%%%%%%%%%%%%%%%%%%%%%%%%%
 Carbon-based nanoelectronics is characterized by relatively small bandgaps,
  and symmetric dispersion relations for electron and holes, associated to a relatively 
  small effective mass. 
  This is the ideal situation for TFETs, whose carbon implementation therefore represents 
  an extremely interesting option~\cite{knoch2008}.
  
  Indeed, a TFET is a gated $p-i-n$ diode where the gate voltage modulates the position of 
  energy bands in the intrinsic channel in order to control interband tunneling between
  the $p$-doped source and the $n$-doped drain.
  When the device is ON, current is dominated by interband tunneling,
  which instead is inhibited when the device is OFF.
  Small energy gaps and small effective masses
  can allow to achieve large ON current, comparable to that of mundane semiconductor field-effect transistors.
  
  An impressive advantage of TFETs is the steep dependence of the current
  on the gate voltage, which can yield a Subthreshold Swing (SS) much smaller than $60$~mV/decade, the minimum
  value achievable at room temperature with FETs.
  The very small SS also allows to implement adequate switches with very small supply voltages, 
  and therefore to operate at extremely low power.
  
  Finally, carbon channels very often have the same dispersion relations and mobility characteristics
  for electrons and holes, which is desirable for the optimization of
  complementary logic gates in terms of switching speed and power consumption~\cite{appenzeller2004}.
  
  TFETs based on carbon nanotubes~\cite{knoch2008}, graphene nanoribbons~\cite{zhang2008,grassi2009} and
  bilayer graphene~\cite{fiori2009,cheli2009a} have been addressed theoretically,
  with simulations and analytical models. 
  Carbon nanotubes and graphene nanoribbons are interesting in terms 
  of the achievable performance, but device operation is extremely sensitive to the bandgap, which in turn 
  strongly depends on the exact number of carbon atoms per ring or along the ribbon width, that is hardly 
  controllable. 
  Bilayer graphene is a two-dimensional channel which does not require very advanced lithography
  and provides a gap tunable with the vertical electric field, up to
  0.15-0.2 eV. 
  Satisfactory device operation is achieved
  with a supply voltage $V_{DD}$ as small as 0.1~V~\cite{fiori2009}.
  
  Very interestingly, epitaxial graphene grown on the Si-terminated face of a SiC substrate is a two-dimensional material 
  exhibiting a bandgap of
  $0.26$~eV~\cite{zhou2007,zhou2008,kim2008} and promising mobility~\cite{tedesco2009}. 
  Such semiconducting gap can be adequate to TFETs suitable for large-scale
  integrated circuits, but is not sufficient for FETs~\cite{cheli2009}.
  %%%%%%%%%%%%%%%%%%%%%%%%%%%%%%%%
  \begin{figure}[bp]
    \centering	
    \includegraphics[width=8cm]{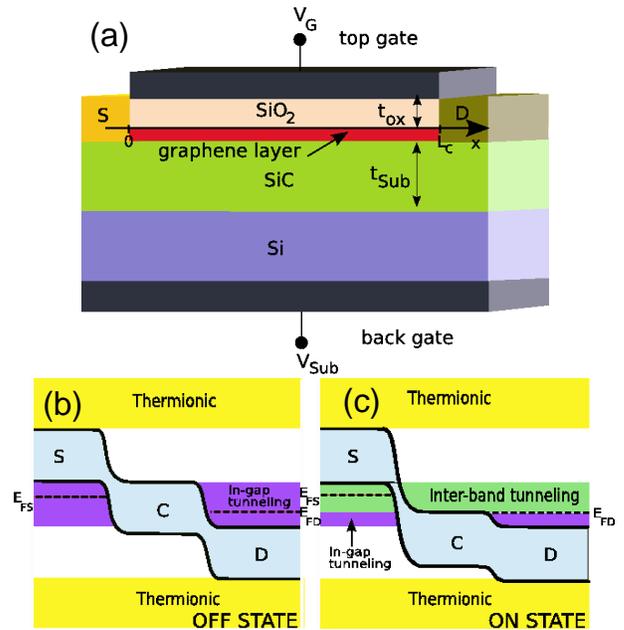}
    \caption{
      (a) Sketch of a TFET device based on graphene on SiC.
      (b) Band profile along the transport direction in the OFF state for a n-channel TFET, and
      ON state (c) of the TFET. 
    }
    \label{fig:1}
  \end{figure}
  %%%%%%%%%%%%%%%%%%%%%%%%%%%%%%%%

  Here we consider the TFET illustrated in 
  \ref{fig:1}(a), in which the channel consists of
  a single graphene layer grown on a SiC substrate of overall thickness
  $t_{Sub}=100$~nm. The top gate is separated by a silicon oxide 
  layer $t_{ox}$ of $1$~nm and is at voltage $V_G$.
  %
  %The TFET is an ambipolar device: If we consider the branch of the I-V
  %characteristics in which it behaves as a $n$-channel devices, the source is 
  %$p$-doped and the drain is $n$-doped.
  %
  Following the evanescent mode approach~\cite{oh2000}, we can estimate
  the natural variation length $\lambda$ of 
  the electrostatic potential in the
  channel as $\lambda = \frac{2}{\pi}(t_{ox}+t_{ch})$, where $t_{ch}=1.1$~nm is the effective distance between the graphene layer
  and the ideal interface of the top dielectric layer~\cite{novoselov2004}.
  Here, we have assumed the evanescent part of the potential to be dominated by the vicinity of the top gate, neglecting the backgate, and, second, we have assumed the same effective dielectric constant for both oxide and graphene regions.
  Because the gate length $L_c$ is always much
  larger than $\lambda$, a long-channel
  behavior is expected.
  Therefore we write $\phi(x)$, the electrostatic potential on the graphene
  channel, in the form 
  \begin{equation}
    q\phi(x)= qA_S e^{-x/\lambda} + qA_D e^{(x-L_c)/\lambda}+\phi_c,
  \end{equation}
  where $\phi_c$ is the potential in the channel away from the contacts,
  $q$ the absolute electron charge,
  $A_S \equiv E_c(0)-E_c(L_c/2)$ and $A_D \equiv E_c(L_c)-E_c(L_c/2)$
  respectively, where $E_c(x)$ represents the conduction band edge 
  in the longitudinal direction $x$.

 For graphene on SiC we adopt the tight-binding
  Hamiltonian proposed in~\cite{zhou2007}, which accounts for the
  energy dispersion curves
  \begin{equation}
    E_{\vec{k},\pm}^{(x)}=\pm
    \sqrt{m^2+t^2|f(\vec{k})|^2}-q\phi(x)
    \label{eq:dispE}
  \end{equation}
  of conduction ($+$) and valence ($-$) bands, whose edges
  respectively give $E_c(x)$ and $E_v(x)$.
 Where $m=0.13$~eV is an effective parameter providing an energy gap of $2m$, $t=2.7$~eV is the nearest-neighbor hopping energy for graphene and 
  $f(\vec{k})$ is the well known off-diagonal element of the $2 \times 2$ graphene $p_z$ -Hamiltonian.
  We assume the potential to be sufficiently
  smooth to rigidly shift graphene band edges. 
  The band edge profiles for the device under investigation in the OFF state
  and in the ON state are shown in Fig.~\ref{fig:1}(b) and \ref{fig:1}(c). 
  
  The device current $I$ is the sum 
  of three major contributions: thermionic, 
  interband tunneling and gap tunneling (due to tunneling through the 
  channel in the gap spectral region).
  Thermionic
  and gap tunneling currents degrade transistor performance, dominating 
  the current when the device is in the OFF state, and therefore must be minimized. 
  The channel current per unit width is
  \begin{equation}
    I = \frac{q}{4\pi^2} \sum_{\pm} \int_{BZ} d\vec{k} T v_x \left[ f(E_{\vec{k},\pm}^{(x_c)}-\mu_S) - f(E_{\vec{k},\pm}^{(x_c)}-\mu_D) \right], 
  \end{equation}
  where the integration runs over the Brillouin zone for conduction
  and valence bands $E_{\vec{k},\pm}^{(x_c)}$, calculated at the
  center of the channel (at $x=x_c=L_c/2$).
  $f$ is the Fermi-Dirac distribution
  function, $v_x$ is the component of the group velocity along the channel. 
  The tunneling coefficient $T$ is calculated via 
  Wentzel-Kramers-Brilluoin approximation in the tunneling
  regions at drain, at source or in the gap, as shown in Fig.~\ref{fig:1}(b).
  In tunneling regions, for a fixed energy $E$ of the incoming particle, we solve the dispersion
  curve (\ref{eq:dispE}) for the complex wavevector describing the evanescent
  mode.
  The integration of the evanescent mode path through the
  barrier gives $T$, as described in more detail in Ref.~\cite{cheli2009}.
  %%%%%%%%%%%%%%%%%%%%%%%%%%%%%%%%
  \begin{figure}[tbp]
    \centering	
    \includegraphics[width=8.5cm]{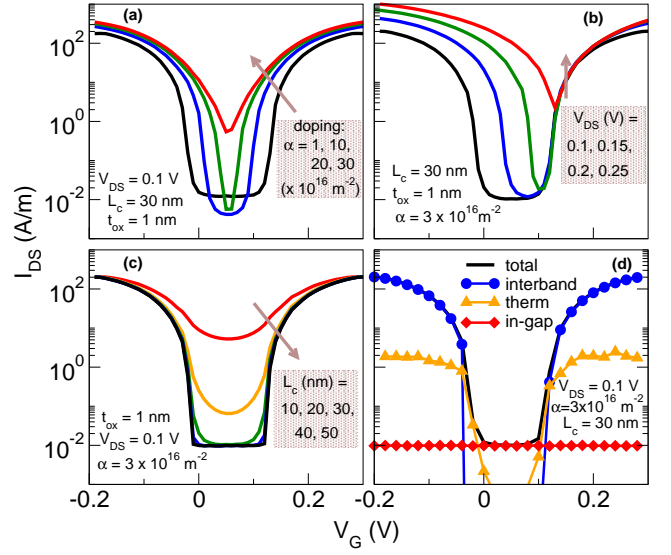}
    \caption{
      Transfer characteristics of a graphene-on-SiC TFET calculated
      for different parameters of the model.
      We analyzed the effects of a variation of doping concentration
      $\alpha$ in (a), of the supply voltage $V_{\rm DS}$ in (b), of the
      channel length $L_c$ in (c).
      In (d) we show an example as the TFET total current is due to
      the summing up of interband tunneling, in-gap tunneling and thermionic current contributions.
    }
    \label{fig:2}
  \end{figure}
  %%%%%%%%%%%%%%%%%%%%%%%%%%%%%%%%   

  In order to compute the channel potential we need to
  self-consistently solve the vertical electrostatics:
    \begin{eqnarray}
    Q &=& -C_G(V_G-\Delta_{G}-\phi_c) -C_{Sub}(V_{Sub}-\Delta_{Sub}-\phi_c)\nonumber\\
    Q &=& Q_h + Q_e,
    \label{eq:VE}
  \end{eqnarray}
  where $C_G=\epsilon_{ox}/t_{ox}$, $C_{Sub}=\epsilon_{SiC}/t_{Sub}$ are the top gate and the
  backgate capacitance per unit area, respectively, whereas $\Delta_{G}$ and $\Delta_{Sub}$ are the flat band voltages of the top and back gates (set here to $0$~V). 
  The charge per unit area in the channel, due to holes and electrons
  ($Q_h$ and $Q_e$), can be also obtained by summing over their
  steady-state distribution in conduction band
  \begin{equation}
    Q_e = -\frac{q}{4\pi^2} \int_{BZ} d\vec{k} [(2-T)f(E_{\vec{k},+}^{(x_c)}-\mu_S)+Tf(E_{\vec{k},+}^{(x_c)}-\mu_D)],  
  \end{equation}
  and valence band  
   \begin{equation}
    Q_h = \frac{q}{4\pi^2} \int_{BZ} d\vec{k} [Tf(\mu_S-E_{\vec{k},-}^{(x_c)})+(2-T)f(\mu_D-E_{\vec{k},-}^{(x_c)})],  
  \end{equation} 
   where we included the possible tunneling barrier at source, drain
   or in-gap tunneling.

  Let us assume the same doping density for source and drain.
  We can therefore define the energy difference between
  conduction(valence) edge and source(drain) Fermi energy at
  source(drain) $\delta=E_c(0)-\mu_S= -\mu_D-E_v(L_c)$.
  It is clear from Fig.~\ref{fig:1}(b,c) that it is possible to shut off the device only if the
  drain-to-source voltage satisfies $|V_{DS}|<(E_{gap}-2\delta)/q$,
  which in the limit of weak doping becomes $|V_{DS}| < 0.26$~V.
  
  Ideally, when the previous condition is satisfied, minimum device current
  is obtained when the channel is intrinsic at
  $L_c/2$, and therefore has zero net charge.
  Accordingly the channel mid gap is $-q\phi_c=(\mu_S+\mu_D)/2$ and
  from (\ref{eq:VE}) we find the  OFF-state gate voltage $V_{\rm OFF}=V_{DS}/2$. 
  The OFF current $I_{\rm OFF}$ is defined as the current obtained 
  with $V_{GS} = V_{\rm OFF}$ and $V_{DS} = V_{DD}$. The ON current
  $I_{\rm ON}$ is the current obtained with $V_{GS} = V_{\rm OFF} + V_{DD}$,
  and $V_{DS} = V_{DD}$.

    In Fig.~\ref{fig:2} we show the TFET transfer characteristics calculated
  for different values of the main parameters of the model: the doping
  fraction $\alpha$ (which is a monotonous function of $\delta$),
  $V_{\rm DS}$, the channel length $L_c$.
  In Fig.~\ref{fig:2}(a) we show the effect of doping concentration
  $\alpha= 10^{16}$, $10^{17}$, $2\times10^{17}$, $3\times10^{17}$~m$^{-2}$,
  corresponding respectively to $\delta\approx 5$, $40$, $65$, $90$~meV.
  
  As expected, the increase of $\delta$ leads to a thinning of the
  range of the OFF-state plateau, 
  in which current is essentially reduced to the thermionic contribution.
  An initial increase of $\delta$ pushes down the source conduction band edge with
  respect to $\mu_S$, and up the drain valence band with respect to
  $\mu_D$, reducing the thermionic hole and electron currents
  respectively.
  For larger $\alpha$ (and $\delta$), one cannot fully shut the channel off 
  because $|V_{DS}| + 2\delta/q$ becomes larger than $E_{gap}$.
 %%%%%%%%%%%%%%%%%%%%%%%%%%%%%%%%
  \begin{figure}[tbp]
    \centering	
    \includegraphics[width=7.3cm]{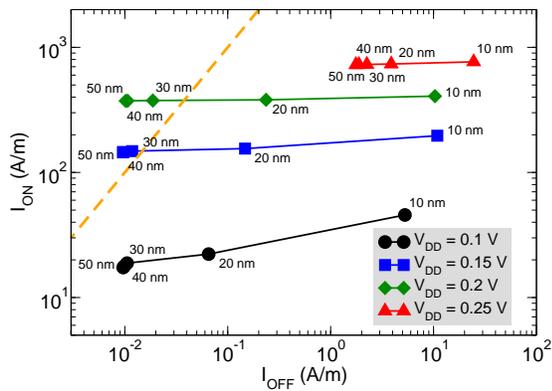}
    \caption{
      $I_{\rm ON}$ and $I_{\rm OFF}$ table for TFET devices with
      $V_{\rm DD}=
      0.1$, $0.15$, $0.2$, $0.25$~V, and $L_c$ between $10$ and
      $50$~nm.
      We consider here $t_{ox}=1$~nm and
      $\alpha=3\times10^{16}$~m$^{-2}$.
      A dashed line marks the field where $I_{\rm ON}/I_{\rm OFF}\ge 10^4$.   
    }
    \label{fig:3}
  \end{figure}
  %%%%%%%%%%%%%%%%%%%%%%%%%%%%%%%% 

  In Fig.~\ref{fig:2}(b) we analyze the operation of the TFET for
  different values of the source-drain voltage $V_{\rm DS}$.
  The OFF-state plateau is reduced with increasing $V_{\rm DS}$, $I_{\rm OFF}$ is constant because limited to 
  the thermionic contribution, whereas larger $I_{\rm ON}$ is obtained.
  For $V_{\rm DS}\approx 0.25$~V and above it is not possible to
  shut the TFET off anymore.
  
  In view of transistor scaling we analyze the effect of a variation of the
  channel length $L_c$ (Fig~\ref{fig:2}(c)).
  We show that for $L_c \ge 30$~nm the OFF current does not depend on
  $L_c$ and is determined by the thermionic current.
  However, a further reduction of $L_c$ leads to an increase $I_{\rm OFF}$,
  due to an increased transparency of the in-gap tunneling barrier.
  The oxide thickness affects tunneling through the contacts
  ($\lambda\propto t_{ox}$), with the main effect of exponentially reducing the tunneling
  $\log{T}\propto -t_{ox}$.
  In Fig.~\ref{fig:2}(d) we show an example of the three current
  contributions occurring in the device. 
  In ON state the tunneling current dominates the device, while it is
  suddenly suppressed going towards the OFF state, where thermionic
  current contribuites to the subthreshold swing and evantually in-gap
  tunneling determines the OFF state current plateau.

  An important performance measure of a transistor is the obtainable
  $I_{\rm ON}/I_{\rm OFF}$ ratio, therefore in Fig.~\ref{fig:3} we plot $I_{\rm ON}$ against
  $I_{\rm OFF}$.
  As noted also before, to shut the device off and have a sufficent
  $I_{\rm OFF}$ dynamics it is required $V_{DS}\le 0.2$~V, togheter
  with sufficiently long channel ($L_c\ge30$~nm) in order to limit in-gap tunneling.
  On the other hand the $I_{\rm ON}$ increases with $V_{DD}$.
  Therefore the best performance is obtained for $V_{DS}=V_{DD}=0.2$~V.  
  Indeed we note that an $I_{\rm ON}/I_{\rm OFF}$ ratio exceeding $10^4$,
  considered adequate for low power operation~\cite{ITRS}, can be
  obtained with a supply voltage of $0.15$ and $0.2$~V, and with a channel length $L_c$
  not smaller than 30~nm.

\begin{figure}[htbp]
    \centering
    \vspace{0.2cm}
    \includegraphics[width=7.5cm]{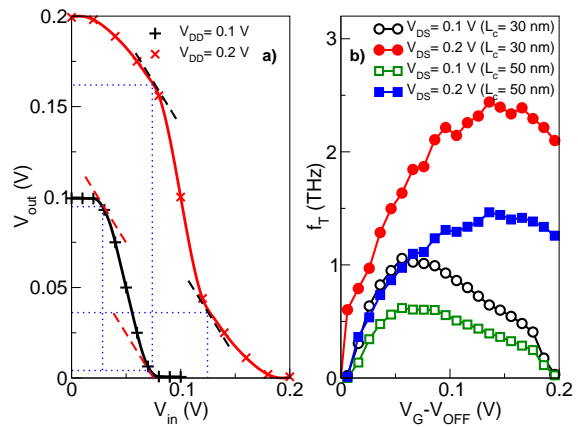}
    \caption{
      (a) Voltage transfer curve for a CMOS
      inverter based on the present device and operating at $V_{DS}=
      0.1$ and $0.2$~V. 
      Dashed lines highlight the points of the curve in which the slope is equal to unity. 
      (b) Cutoff frequency calculated for TFETs with $V_{DS}=
      0.1$ and $0.2$~V, for $L_c=30$ and $50$~nm.
    }
    \label{fig:4}
  \end{figure}

  Now we analyze the performance of our device for digital applications.
  In Fig.~\ref{fig:4}(a) we plot the transfer characteristics of a CMOS inverter 
  with identical 
  $n$ and $p$-type TFETs, with the channel of graphene on SiC with
  $L_c=30$~nm and $t_{ox}= 1$~nm, for a supply voltage of $0.1$~V and
  $0.2$~V.
  We extract the noise margins (NMs) for the CMOS inverter, which are a
  measure of logic gate robustness to disturbs and noise, as ${\rm NM} =22$~mV and
  $37$~mV, respectively for $V_{DD}=0.1$ and $0.2$~V. NMs are not extremely good,
  but in line with what one could expect, given the low supply voltage.
   %%%%%%%%%%%%%%%%%%%%%%%%%%%%%%%%
  %%%%%%%%%%%%%%%%%%%%%%%%%%%%%%%%

  We also performed an analysis of transistor intrinsic delay and 
  transition frequency $f_T$ of the considered TFET.
  We consider operating voltages of $0.1$~V and $0.2$~V and a channel
  length of $30$ and $50$~nm.
  In particular in Fig.~\ref{fig:4}(b) we show the cutoff frequency as a function of the overdrive,
  calculated in the quasi-static approximation~\cite{burke2004}
  $f_T= g_m /(2 \pi C_G L_c)$, where
  $g_m \equiv \partial I_{DS}/ \partial V_{GS}$ is the
  transconductance, and $C_G \equiv \partial Q/\partial V_{GS}$ evaluated for
  $V_{GS} = V_{DD}+V_{\rm OFF}$, is the capacitance seen by the gate
  contact.
  The intrinsic delay time, which represents a figure of merit for digital
  application, is $\tau
  \equiv V_{DD} C_{G}/I_{\rm ON} $.
  For a supply voltage of $0.1$~V, we obtain $\tau= 0.33$~ps and
  $0.54$~ps, respectively for a channel of length $30$ and $50$~nm.
  With a supply voltage of $0.2$~V, $\tau= 0.12$~ps for $L_c = 30$~nm
  and $\tau= 0.22$~ps for $L_c = 50$~nm.

 In conclusion, we examined the maximum achievable performance of a 
  TFET based on  epitaxial graphene on SiC substrate, by means of an accurate
  semi-analytical model.
  We explored the parameter space analyzing its
  performance under variations of the doping level, supply
  voltage, channel length and oxide thickness.
  We show that an $I_{\rm ON}/I_{\rm OFF}$ exceeding $10^4$ can be easily
  achieved by choosing adequate geometry and bias, and a channel length not smaller than 30~nm,
  even for a supply voltage as small as 0.15~V. Noise Margins of a minimum size inverter are reasonable but not impressive.
  Intrinsic delay times and transition frequency are very promising, especially considering the very low supply voltage, 
  and make the proposed TFET a strong candidate for extreme low-power carbon nanoelectronics.

%%%%%%%%%%%%%%%%%%%%%%%%%%%%%%%%%%%%%%%%%%%%%%%%%%%%%%%%%%%%%%%%%%%%
%%%%%%%%%%%%%%%%%%%%%%%%%%%%%%%%%%%%%%%%%%%%%%%%%%%%%%%%%%%%%%%%%%%%

%\bibliographystyle{IEEEtran}
%\bibliography{IEEEabrv,bibf}
%\bibliography{bibf}

\end{document}